\documentclass[pra,onecolumn,10pt,aps,longbibliography,notitlepage]{revtex4-1}
\usepackage[dvips]{graphics,graphicx}
\usepackage[utf8]{inputenc}
\usepackage[T1]{fontenc}
\usepackage{amsmath}

\usepackage{amssymb}
\usepackage{epstopdf}
\usepackage[usenames, dvipsnames]{color}
\usepackage{graphicx}

\usepackage{amsfonts}
\usepackage{eufrak}
\usepackage{mathrsfs}
\usepackage{bm}
\newcommand{\scs}{\scriptscriptstyle}

\begin{document}

\title{Enhanced sensitivity of an all-dielectric refractive index sensor with optical bound state in
the continuum}
\author{Dmitrii N. Maksimov$^{1,2}$, Valery S. Gerasimov$^{1,3}$, Andrey A. Bogdanov$^4$,
Sergey P. Polyutov$^1$}
\affiliation{$^1$International Research Center of Spectroscopy and Quantum Chemistry, Siberian Federal University,
 660041, Krasnoyarsk, Russia}
\affiliation{$^2$Kirensky Institute of Physics, Federal Research Center KSC SB RAS, 660036, Krasnoyarsk, Russia}
\affiliation{$^3$Institute of Computational Modeling SB RAS, Krasnoyarsk, 660036, Russia}
\affiliation{$^4$Department of Physics and Engineering, ITMO University, 191002, St. Petersburg, Russia}
\begin{abstract}
The sensitivity of a refractive index sensor based on an optical bound state in the continuum is considered. Applying Zel'dovich perturbation theory we derived an
analytic expression for bulk sensitivity of an all-dielectic sensor utilizing symmetry protected in-$\Gamma$ optical bound states in a dielectric grating. The upper sensitivity limit is obtained. A recipe is proposed for obtaining the upper sensitivity limit by optimizing the design of the grating. The results are confirmed through direct numerical simulations.
\end{abstract}

\maketitle

\section*{Introduction}

Recently, we have witnessed a surge of interest to bound states
in the continuum (BICs) \cite{Hsu16, Koshelev19, sadreev2021interference} that have revolutionized nanophotonics by paving a way to high throughput optical sensing devices with enhanced light-matter interaction at the nanoscale \cite{Molina2012, Romano18, BS1, Penzo2017, Zhen2014, Mocella2015, koshelev2019meta, Zito2019}. The most remarkable feature of BICs is the occurrence of high-quality Fano resonances
in the transmittance spectrum \cite{Shipman,SBR,Blanchard16,Bulgakov18b}.
The Fano resonances emerge when the system's symmetry is broken, so the otherwise localized BIC mode can couple with impinging light \cite{koshelev2018asymmetric, Maksimov20}. The spectral position of these extremely narrow Fano resonances is affected by the refractive index of the surrounding medium allowing to design optical sensors with an excellent figure of merit ({FOM}) \cite{Liu17,Romano18b, Romano18a, Yesilkoy19, Mukherjee19, Vyas20} as the narrow Fano feature can be easily resolved in the spectral measurements.
Despite the unsurpassed FOM, the major drawback of the dielectric sensors in comparison against the plasmonic ones is a noticeably (approximately five times) less sensitivity \cite{Bosio19}. Nowadays, comparative analysis
of dielectric sensors performance is already available in the existing literature \cite{Pitruzzello18, Wang21}. Yet, to the best of our knowledge, there is no exhaustive theory providing
a clear optimization procedure that would lead to the maximal sensitivity of a BIC sensor. In our previous paper \cite{Maksimov20} we argued
that the maximal sensitivity of a BIC sensor is given by
$S_{\mathrm {max}}=\lambda_{\mathrm{BIC}}/n_c$, where $\lambda_{\mathrm{BIC}}$ is the vacuum wavelength of the BIC, and
$n_c$ is the the cladding fluid refractive index, i.e. the physical quantity whose variation is  measured by the sensor.
In this paper we provide a rout for achieving the above limit
by considering an optical BICs whose spectral position approach
the frequency cut-offs of the first order diffraction channels.
An analytic expression for a BIC vacuum wavelength shift is
derived by applying the Zel'dovich perturbation theory \cite{zel1961theory}. The primary advantage of the Zel'dovich approach is that it can be applied to optical delocalized eigenmodes\cite{Lai90} to have already been proved useful in theory of plasmonic sensors \cite{Zalyubovskiy12}. The onset of diffraction usually has
a negative impact on high-Q resonances, particularly on BICs which are typically destroyed by the radiation losses \cite{Sadrieva17}. However, in the situation when the first order diffraction cut-off frequency is not yet exceeded the evanescent field are demonstrated to provide the largest overlap with the cladding fluid leading to enhanced sensitivity \cite{Romano19}.


\section*{The system}

\begin{figure}[t]
\centering
\includegraphics[width=0.9\textwidth,height=0.25\textheight,
,trim={1cm 11.2cm 1.0cm 11.5cm},clip]{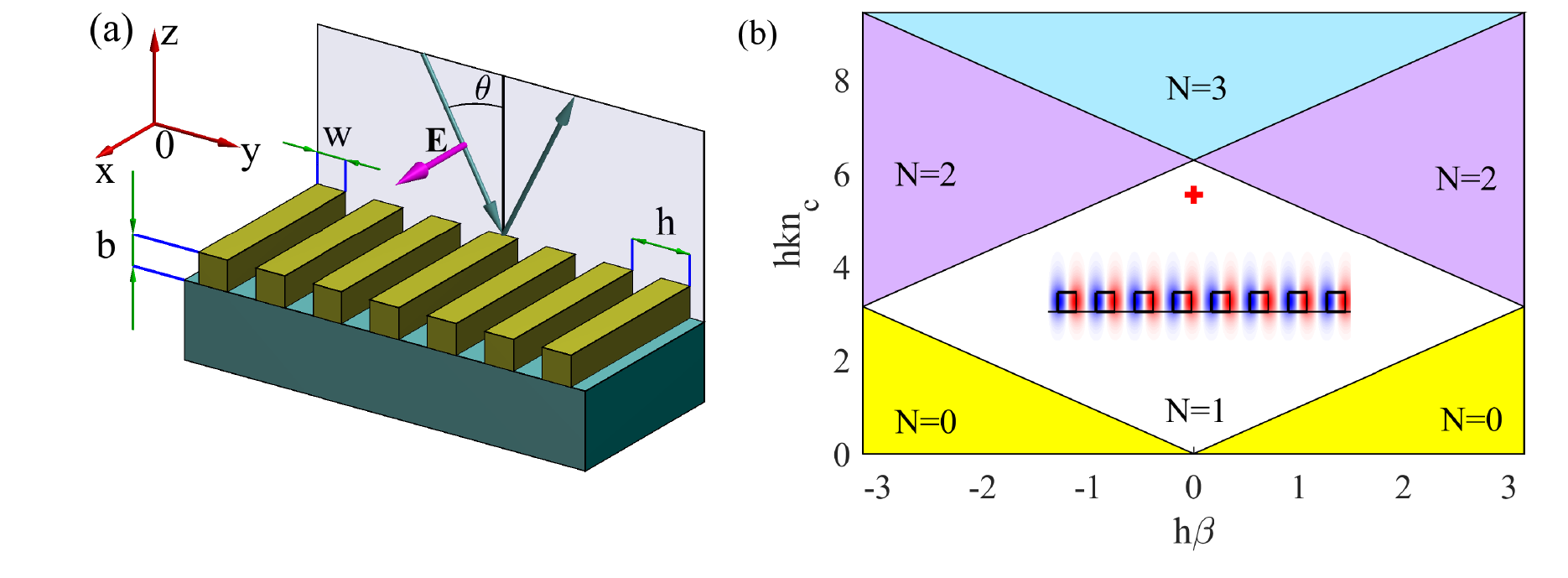}
\caption{BIC in dielectric grating. (a) ${\rm TiO}_2$ dielectric grating on a glass substrate,
$h=300nm$, $b=0.5h$, $w=0.5h$, $n_g=2.485$, and $n_s=1.5$. The magenta arrow is the electric vector of an incident $s$-wave; (b) Number of scattering channels allowed for diffraction.
The red star indicates the position of a symmetry protected BIC with $k_{{\rm BIC}}=3.262$, $n_c=1.7$. The BIC mode profile is depicted below its spectral position in procedure defined units with red areas corresponding to positive $E_x$ whereas in the blue areas $E_x$ is negative.}
\label{fig1}
\end{figure}

One of the most important platforms for implementing optical BICs is subwavelength dielectric gratings \cite{Marinica08, Sadrieva17, Monticone17, Bulgakov18, Bulgakov18a, Lee19, Bykov19, Gao19, Hemmati19} which have been extensively studied for sensing applications
with both dielectric \cite{Maksimov20, zhang2021high, Finco21, Park21, Mesli21} and plasmonic structures \cite{Jia21}.
The system under consideration is a subwavelength dielectric grating of period
$h=300{\rm nm}$ made of ${\rm TiO}_2$. In our numerical simulations we use $n_g=2.485$ as the refractive index of the grating while the material
losses are ignored so the absorption coefficient is set to zero. The ${\rm TiO}_2$ grating is placed on top of a glass substrate with refractive index $n_s=1.5$ as shown in Fig.~\ref{fig1}~(a).
The geometric parameters of the grating are specified in the caption to Fig.~\ref{fig1}.
Propagation of TE modes is controlled by the Helmholtz equation
\begin{equation}\label{basic}
    \nabla^2 E_x(y,z)+k^2\epsilon(y,z) E_x(y,z)=0,
\end{equation}
where $E_x$ is the $x$-component of the electric field, $\epsilon=f(y,z)$~--- the coordinate dependant dielectric function, $\nabla^2$~--- the 2D Laplacian
operator, and
$k=\omega/c$ is the vacuum wave-number.
The spectrum of far-field diffraction channels available for scattering
is given by
\begin{equation}\label{Wood}
    n_ck^2=\left(\beta+\frac{2\pi m}{h}\right)^2+k_z^2, \ m=-\infty, \ldots, -1, 0, 1, \ldots, \infty,
\end{equation}
where $k_z$ is the $z$ component of the outgoing wave vector, $\beta$ is the propagation constant with respect to the $y$-axis, and $n_c$ is the refractive
index of the cladding fluid. All simulation results referenced throughout the paper have been obtained with application of FDTD Lumerical photonics simulation software solution.

Following our previous work \cite{Maksimov20} we consider a symmetry protected in-$\Gamma$ BIC
which does not radiate to the far field because it is symmetrically  mismatched with the zeroth-order
diffraction channel at strict normal incidence. In engineering a symmetry protected BIC it is of
key importance to ensure that the higher diffraction orders are not allowed at the BIC wavelength.
Once diffraction is allowed either in the substrate or in the superstrate (cladding fluid) the BIC is
destroyed being transformed into a leaky mode radiating into the diffraction channels \cite{Sadrieva17}. The onset of
diffraction is typically detected by occurrence of Wood-Rayleigh anomaly in the scattering spectra~\cite{Wood1902,Rayleigh1907}.
The positions of
Wood's anomalies are given by the following equations
\begin{align}\label{Wood2}
   & n_ck=|\beta|, \nonumber \\
   & n_ck=\beta+\frac{2\pi m}{h} \ \mathrm{if} \ m>0 \nonumber, \\
   & n_ck=-\beta-\frac{2\pi m}{h}\ \mathrm{if} \ m<0.
\end{align}
In Fig.~\ref{fig1} (b) we demonstrate the number of diffraction channels $N$ allowed in the cladding
depending on the spectral parameters of the incident wave, namely its vacuum wavelength, $k$ and the propagation constant in the $y$-direction, $\beta$.  Each colored domain in Fig.~\ref{fig1}~(b) corresponds to the specified value  of $N$. Wood's anomalies occur when the boundaries between the domains with $N>1$ are
crossed under variation of the spectral parameters of the incident wave.
The symmetry protected BICs can be found in the domain of specular reflection $N=1$. The spectral position of such a BIC with $n_c=1.7$ is shown in \ref{fig1}~(b) by a red plus sign. The BIC mode profile is set in the middle. It is important for our future purpose to point out that the  BIC mode profile has the following asymptotic far-field expression
\begin{equation}\label{far}
    E_x\propto e^{-\varkappa z}\sin\left(\frac{2\pi y}{h}\right),
\end{equation}
where
\begin{equation}\label{kappa}
    \varkappa=\sqrt{\left(\frac{2\pi}{h}\right)^2-\epsilon^{\scs{(0)}}_{\infty}k^2}
\end{equation}
with $\epsilon^{\scs{(0)}}_{\infty}=n_c^2$ and $y=0$ corresponding to the geometric center of the unit cell.
Notice that in Equation~\eqref{far} we only retained the contribution
of the first order evanescent diffraction channels, since the higher
order channels decay much faster with the increase of $z$.

\section*{Perturbation theory}

The Zel'dovich \cite{zel1961theory} perturbation approach is introduced by writing
a series expansion in the increment of the dielectric function
$\Delta\epsilon$
\begin{align}\label{perturb}
   & \epsilon=\epsilon^{\scs{(0)}}+\Delta\epsilon, \nonumber \\
   & k=k^{\scs{(0)}}+\Delta k, \nonumber \\
   & E_x=E_x^{\scs{(0)}}+E_x^{\scs{(1)}}.
\end{align}
By definition the function $\epsilon^{\scs{(0)}}$ is the total coordinate dependent dielectric function with the reference value of the cladding
fluid refractive index, whereas $\Delta\epsilon$ is the increment of the
dielectric constant of the cladding fluid.
Substituting the above into Equation~\eqref{basic} we have
up to the first perturbation order
\begin{align}\label{first}
    & \nabla^2E_x^{\scs{(0)}}+\epsilon^{\scs{(0)}}(k^{\scs{(0)}})^2E_x^{\scs{(0)}}=0, \\ \label{second}
    & \nabla^2E_x^{\scs{(1)}}+\epsilon^{\scs{(0)}}(k^{\scs{(0)}})^2E_x^{\scs{(1)}}=-\Delta\epsilon (k^{\scs{(0)}})^2E_x^{\scs{(0)}}-
      2\epsilon^{\scs{(0)}} \Delta kk^{\scs{(0)}}E_x^{\scs{(0)}}.
\end{align}
Following Zel'dovich we multiply Equation~\eqref{second} by
$E_x^{\scs{(0)}}$ and integrate within the limits specified below
\begin{equation}\label{int}
\int\limits_{-{h}/{2}}^{{h}/{2}}dy
\int\limits_{-d}^{d}dz
E_x^{\scs{(0)}}[\nabla^2E_x^{\scs{(1)}}+\epsilon^{\scs{(0)}}(k^{\scs{(0)}})^2E_x^{\scs{(1)}}]=-\int\limits_{-{h}/{2}}^{{h}/{2}}dy
\int\limits_{-d}^{d}dz[\Delta\epsilon (k^{\scs{(0)}})^2+
      2\epsilon^{\scs{(0)}} \Delta kk^{\scs{(0)}}](E_x^{\scs{(0)}})^2,
\end{equation}
where $d>b$ is an arbitrary distance from the grating.
By applying Green's theorem together with Equation~\eqref{first} one rewrites
the L.H.P. of Equation~\eqref{int} in the following manner
\begin{equation}\label{int2}
\int\limits_{-{h}/{2}}^{{h}/{2}}dy
\int\limits_{-d}^{d}dz
E_x^{\scs{(0)}}[\nabla^2E_x^{\scs{(1)}}+\epsilon^{\scs{(0)}}(k^{\scs{(0)}})^2E_x^{\scs{(1)}}]=\int\limits_{\ell} d\ell
\left(E_x^{\scs{(0)}}\frac{\partial E_x^{\scs{(1)}}}{\partial n} -E_x^{\scs{(1)}}
\frac{\partial E_x^{\scs{(0)}}}{\partial n}\right),
\end{equation}
where $\ell$ is a path encircling the integration domain of
the L.H.P. in the clockwise direction. The integrals along the lines
$y=-h/2$ and $y=h/2$ cancel each other due to periodicity while the
integral along $z=-d$ can be neglected as we assume that the BIC field decays faster in
the substrate because of a lesser refractive index. This leads us to
\begin{equation}\label{int3}
\int\limits_{-{h}/{2}}^{{h}/{2}}dy
\int\limits_{-d}^{d}dz
E_x^{\scs{(0)}}[\nabla^2E_x^{\scs{(1)}}+\epsilon^{\scs{(0)}}(k^{\scs{(0)}})^2E_x^{\scs{(1)}}]=\int\limits_{-h/2}^{h/2} dy
\left(E_x^{\scs{(0)}}(y,d)\frac{\partial E_x^{\scs{(1)}}(y,d)}{\partial z} -E_x^{\scs{(1)}}(y,d)
\frac{\partial E_x^{\scs{(0)}}(y,d)}{\partial z}\right).
\end{equation}
Equating the R.H.P. of Equation~\eqref{int3} to that of Equation~\eqref{int}
and inserting the following asymptotic expressions
\begin{align}
    & \frac{\partial E_x^{\scs{(0)}}}{\partial z}=-\varkappa^{\scs{(0)}} E_x^{\scs{(0)}}, \nonumber \\
    & \frac{\partial E_x^{\scs{(1)}}}{\partial z}=
    \frac{1}{2\varkappa^{\scs{(0)}}}[\Delta\epsilon (k^{\scs{(0)}})^2+2\epsilon^{\scs{(0)}}_{\infty}k^{\scs{(0)}}\Delta k]E_x^{\scs{(0)}}
    -\varkappa^{\scs{(0)}} E_x^{\scs{(1)}},
\end{align}
one has
\begin{equation}\label{perturbation_final}
    \frac{\Delta k}{\Delta \epsilon}=-\frac{k^{\scs{(0)}}}{2}I,
\end{equation}
where
\begin{equation}\label{I}
    I=
    \frac{ \displaystyle\int\limits_{S_{\rm c}}dS\left(E_x^{\scs{(0)}}\right)^2+
   \frac{1}{2\varkappa^{\scs{(0)}}}\displaystyle\int\limits_{-h/2}^{h/2} dy \left[E_x^{\scs{(0)}}(y,d)\right]^2}
    {\displaystyle\int\limits_{S_{\rm tot}}dS\epsilon^{\scs{(0)}}\left(E_x^{\scs{(0)}}\right)^2
    +\frac{\epsilon^{\scs{(0)}}_{\infty}}{2\varkappa^{\scs{(0)}}}\int\limits_{-h/2}^{h/2} dy \left[E_x^{\scs{(0)}}(y,d)\right]^2}.
\end{equation}
Thus, for the differential sensitivity $S=\Delta\lambda_{\scs{\mathrm{BIC}}}/\Delta n_c$
we have
\begin{equation}\label{sensitivity}
S=\lambda n_c I.
\end{equation}
On approach to the cut-off of the first order diffraction channels $\varkappa\rightarrow 0$
we have
\begin{equation}\label{Smax}
S=\frac{\lambda_{\scs{\mathrm{BIC}}}}{n_c}
\end{equation}
which coincides with the upper sensitivity limit predicted in \cite{Maksimov20}. Before proceeding to numerical validation of the newfound results some comments are due on the choice of $d$. It is obvious that $d$ should be taken large enough to guarantee that the
far-field asymptotic behaviour be given by Equation~\eqref{far}. In the domain of specular reflection at the normal incidence the evanescent fields of the second order diffraction
channels decay faster than $e^{-2\pi z/h}$. Therefore, for a practical choice it is sufficient to take
$d>\lambda_{\scs{\rm BIC}}$. At the save time, from the computation viewpoint the choice of $d$ determines the size of the computational domain. Technically, $d$ is the distance at which the PML absorbers are placed to set-up reflectionless boundary conditions. Thus, application of Equations (\ref{I},~\ref{sensitivity}) allows us to predict the sensitivity
of BICs with the evanescent fields extended beyond the computational domain.

\section*{Numerical validation}
\begin{figure}[t]
\centering
\includegraphics[width=0.9\textwidth,height=0.22\textheight,
,trim={0cm 11.2cm 0.0cm 11.5cm},clip]{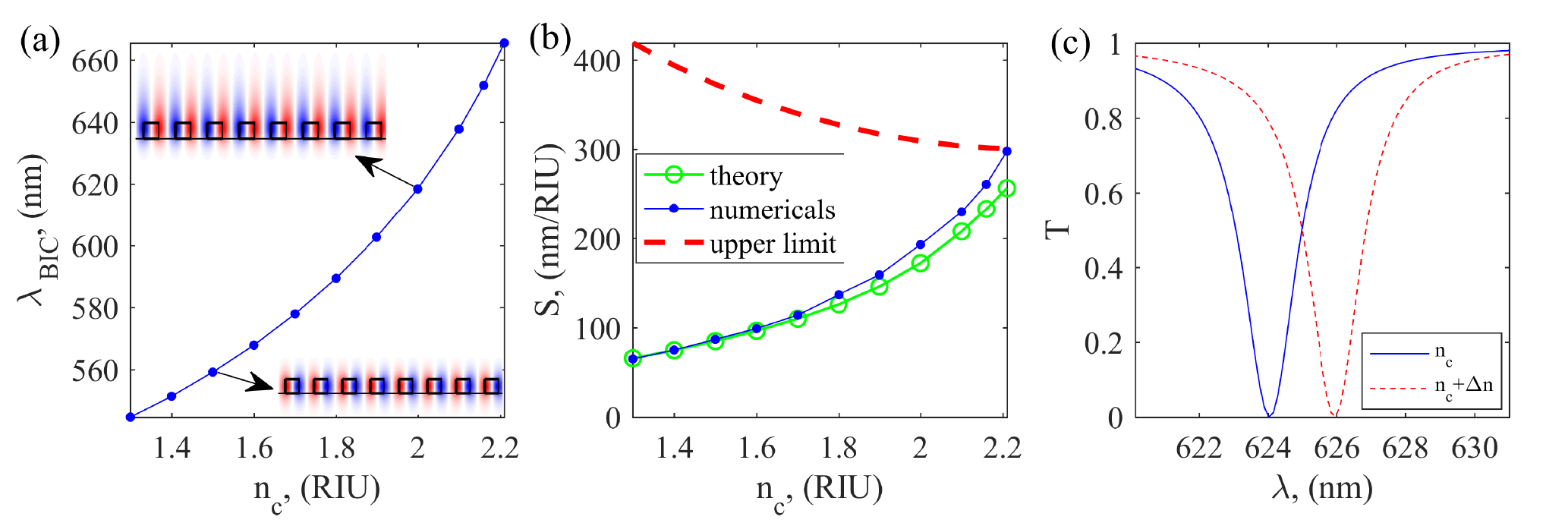}
\caption{Fano resonance shift and differential sensitivity. (a) The vacuum wavelength of the symmetry protected BIC as a function of the cladding fluid refractive index. (b) Differential sensitivity obtained from Equation~\eqref{sensitivity}~--- green circles, and through the shift of the Fano resonance~--- blue dots at $\theta=2 (\mathrm{deg})$. (c) Shift of the Fano resonance obtained at $\theta=2 (\mathrm{deg})$, $n_c=2.0$ with $\Delta n=0.01$. }
\label{fig2}
\end{figure}
In Fig.~\ref{fig2}~(a) we plotted the vacuum wavelength of the symmetry protected BIC at different
values of $n_c$. One can see from Fig.~\ref{fig2}~(a) that $\lambda_{\mathrm{\scs{BIC}}}$ decreases with
the increase of $n_c$ especially at larger values of  $n_c$.
In Fig.~\ref{fig2}~(b) we plotted the values of differential sensitivity obtained by two different
methods. The "theoretical" result is obtained by directly applying Equation~\eqref{sensitivity} meanwhile the "numerical" values are obtained finding the shift of a BIC-induced Fano resonance
under illumination by a plane wave at the incidence angle $\theta=2 (\mathrm{deg})$.
To keep the approach consistent while changing $n_c$, the incidence angle $\theta$ is defined in reference to the wave vectors in the outer space (air), so the incident
angle within the cladding fluid $\theta_c$ is found through the formula
\begin{equation}\label{snell}
n_c\sin(\theta_c)=\sin(\theta).
\end{equation}
The typical picture of the shift in the resonance position is shown in Fig.~\ref{fig2}~(c). In Fig.~\ref{fig2}~(b) we also plotted the maximal value of the sensitivity according to Equation~\eqref{Smax}. One can see from Fig.~\ref{fig2}~(b)
that with the increase of $n_c$ the sensitivity approach the upper limit given by Equation~\eqref{Smax}. Notice that at larger $n_c$ the numerically observed sensitivity deviates
from the theoretical expressions (\ref{I},~\ref{sensitivity}). This is because the first order frequency cut-offs are lower  at $\theta\neq0$ according to Equation~\eqref{Wood2}, see also Fig.~\ref{fig2}~(b). In our case the onset of diffraction at $n_c=2.21$ destroys the high-Q resonance with the chosen angle of incidence $\theta=2 \mathrm{(deg)}$ although
the BIC proper is not yet destroyed. Finally, notice that the observed sensitivity enhancement can be easily understood from Fig.~\ref{fig2}~(a) which is complemented
with subplots of two BIC mode profiles. One can see that at larger $n_c$ the BIC fields
are further extended into the upper half-space providing a better overlap with the cladding fluid.

\section*{Conclusion}

In this paper we have demonstrated an approach allowing to achieve the upper sensitivity limit for an all-dielectric sensor based on an optical bound state in the continuum (BIC).
The analytic expressions for computing the bulk sensitivity from the BIC vacuum wavelength and mode profile have been derived as Equations~(\ref{I},~\ref{sensitivity}). The most remarkable feature of the obtained expressions is that the maximal sensitivity is independent of the material and geometric parameters of the grating allowing for freedom in choosing of the constituent dielectric. All that is necessary for achieving the maximal sensitivity is to vary the geometric parameters for the BIC eigenfrequency approaching
the cut-offs of the first order diffraction channels. The BICs are topologically protected objects \cite{Zhen2014,Bulgakov17,Bulgakov17a} and therefore are bit destroyed under variation of geometric parameters complying with the structure point group symmetry as
far as the BIC eigenfrequency remains in the spectral domain of the specular reflection~\cite{Bulgakov18}. In the regime of maximal sensitivity the vacuum wavelength of the in-$\Gamma$ BIC is always given by
\begin{equation}
    \lambda_{\scs{\mathrm{BIC}}}=n_ch,
\end{equation}
while the maximal sensitivity is simply $S_{\scs{\mathrm{max}}}=\lambda_{\scs{\mathrm{BIC}}}/n_c$. Notice that due to the scaling invariance of Maxwell's equations, $\lambda_{\scs{\mathrm{BIC}}}$ can be tuned to any desired wavelength by isometric transformation of the grating.

In the numerical example proposed the maximal sensitivity has been achieved with
a relatively large value of the cladding fluid refractive index $n_c=2.21$. This, of course, out of the range required in practical applications which are stuck around the refractive index of water. Importantly, achieving the maximal sensitivity requires the
refractive index of the cladding larger than that of substrate. If the situation is the opposite, the first order
diffraction channels will first open in the substrate~\cite{Sadrieva17} destroying the BIC before the maximal overlap with the cladding has been achieved. There are two possible solutions to this problem. The first is, obviously, applying a low index substrate~\cite{Schubert07}. The second is using
a substrate of a properly chosen Bragg reflector~\cite{Bikbaev21} which would isolate the lower half-space
by a photonic band gap for the outgoing waves of the first order of diffraction. This problem will be addressed in the future studies.

\bibliography{BSC_light_trapping}


\section*{Author contributions statement}

D.N.M. and A.A.B conceived the idea presented. D.N.M. derived the analytic results.
V.S.G. ran numerical simulations. S.P.P. supervised the project. All authors
have equally contributed to writing the paper.

\section*{Additional information}

\

\textbf{Competing interests}: The authors declare no competing interests.

\

\textbf{Data availability}:
The data that support the findings of this study are available from the corresponding author, D.N.M., upon reasonable request.



\end{document}